# Evolutionary game of N competing AIMD connections

Oleksii Ignatenko[1], Oleksandr Synetskyi[1]

[1] Institute of Software Systems NAS of Ukraine
`oignat@isofts.kiev.ua`

**Abstract.** This paper deals with modeling of network's dynamic using evolutionary games approach. Today there are many different protocols for data transmission through Internet, providing users with better or worse service. The process of choosing better protocol could be considered as a dynamic game with players (users), trying to maximize their payoffs (e.g. throughput). In this work we presented the model of network's dynamic using differential equations with discontinuous right side and proved existence and uniqueness of solution, formulated payoff matrix for a network game and found conditions of equilibrium existence depending of loss sensitivity parameter. The results are illustrated by simulations.

**Key Terms.** MathematicalModel, MultiAgentSystem, Infrastructure

## 1    Introduction

This work deals with analytical model of competition between data flows in a network. Each flow belongs to a selfish non-cooperative user. Selfishness means that a user wants to utilize as much network resources as possible and the term non-cooperative does not imply that the users do not cooperate, but it rather means that any cooperation must be based on individual information with no communication or coordination among the players.

This competitive situation leads to obvious conflict when summary users' demand is bigger then network supply. If this happens network drops users' data, so generally it is not an event good for users. From the other side, network underloading (when demand is smaller then network capacity) is also undesirable because it leads to losses in efficiency of resource using.

From the beginning the Internet has been regulated by protocol (algorithms controlling user's behavior and implicitly forming network behavior as well) called TCP (the Transmission Control Protocol), introduced in the 1970s to provide reliable data transfer. However, that version was proved to be unreliable because it caused phenomenon known as *congestion collapse*. Later Van Jacobson improved TCP by developing a congestion control mechanism.

The main idea is rather simple rule for a user's behavior, depending on information about network state:

*if network is underloaded – increase your rate,*

*if network is overloaded – decrease your rate.*

Delivering information about network state to the end user is a challenging problem and a crucial part of any feedback based protocol. As a rule a user has knowledge about successful delivery of his data (in other words he knows that network is probably underloaded) and about overload event (if he doesn't receive successful ACK – acknowledgement packet) with some delay. This type of information is called binary feedback. The natural rate control based on this information called AIMD [1] (additive increase, multiplicative decrease) scheme. There are another possibilities, but it was proved that AIMD algorithm will oscillate near the point of effective (all bottlenecks will be loaded) and fair (in some sense) allocation of network resources.

Nowadays, TCP isn't one protocol but a big family (number keeps increasing) of algorithms implementing different implementation of the origin idea. Protocol development went through the competitive evolution between different protocols, abandonment of some protocols and appearance of new ones. The possibility to deploy new versions of protocols gives users control to improve performance of his connection by choosing suitable algorithm. When many users are trying to achieve better performance it is difficult to predict consequences of such a competition. There is a problem how to ensure stable, fair and effective network behavior in the situation of dynamic and antagonistic interaction of selfish users. We address this problem with evolutionary game approach.

First we make short introduction of using game and control theory frameworks related to network problems. The work of F.Kelly et al. [2] was the first example of considering of resource allocation as an optimization problem. Later many authors [see for example 3 – 5] have developed generalizations and variations of this framework. There are many approaches of investigation of complex networks from different directions (static, dynamic, deterministic etc). The evolutionary games concept is a part of game theory that focuses on studying interactions between populations rather than individual players. One of the earliest publications about the use of evolutionary games in networking is [6] that study through simulations some aspects of competition between TCP users. The evolutionary games based on the concept of the ESS (Evolutionary Stable Strategy), defined in 1972 by the biologist Maynard Smith [7]. The ESS concept has been used in [8] in the context of ALOHA with power control. Fundamental survey of applications of game theory to networks is [9]. In this paper we develop the line of research presented in [6] by Altman et al. They considered a model of users which are using two different TCP connections – peaceful and aggressive. For this model it was shown that dynamic of this process described by difference equation has a stable solution and users payoffs are forming a structure of evolutionary game known as Hawk-Dove game. Also there were identified conditions under which equilibrium is evolutionary stable. However, there are limitations of the proposed approach.

First, the proposed method could not be generalized on the case of three or more protocols. Second, the network considered in this paper has very simple topology (a single node) and no extensions in this direction were proposed.

In current work we propose more general approach to evolutionary game of N different AIMD connections competing for resource. We found the solution using fixed point theorem, which makes possible generalization on N connection case and complex network topology. We formulated a game for player population, with strategy of choosing the best protocol. We found conditions for equilibrium existence and described it properties. Also we illustrate conditions by simulations of system dynamic.

## 2  Model

Consider a network of $M$ processing nodes connected in some topology. Every node has at least one service link with limited overall capacity (or processing rate) $p_i$, $i = 1,...,M$. Let $I$, $K$ be a set of nodes indexes $\{1,...,M\}$ and service links indexes $\{1,...,L\}$ respectfully. There are $N$ users, connected to this network. Let $x_j(t)$ be the transmission rate of $j$ users, where $j \in J = \{1,...,N\}$. There is natural assumption about vector of rates $\bar{x} = (x_1,...,x_N)$: $x \in R_+^N$. If sum of transfer rates of data flows using the node's links is equal or bigger than the node capacity then overload event occurs (overload here is a synonym of packet loss). This scheme is an idealized model of widely deployed Droptail scheme. We will assume that routing is deterministic and uncontrolled and information about overload delivers to users momentarily. Let us fix the following notation (used for example in [10]) throughout this paper.

Denote $u_k(t)$, $k \in K$ as the service rate of $k$'s link. The constituency matrix is the $M \times L$ matrix C whose $c_{ij}$ element is equal to 1 if $i$'s link belongs to $j$'s node and otherwise is 0. Using this matrix we define a set $U$ as $\{u \in R_+^K \mid Cu \leq 1\}$. The set $U$ contains all possible service rates for the system. Let $P$ be $diag\{p_1,...,p_M\}$ - diagonal matrix.

The routing matrix $R$ is the $M \times M$ matrix defined for $i, j \in P$. Element $r_{ij}$ is equal to 1 if the output of $i$'s link is the input of $j$'s link and otherwise is 0. The input matrix $A$ is the $L \times N$ matrix defined for $i \in K, j \in J$. Element $a_{ij}$ is equal to 1 if $j$'s user uses $i$'s link and otherwise is 0.

### 2.1  Overload conditions

When the system produces overload and how one can analytically predict it? This is an important problem of network modeling.

**Proposition 2.1.** (Stability condition) If user's vector of rates $x(t)$, $t \in [t_0, t_1]$ satisfy condition $\Xi \bar{x}(t) < 1$, where matrix $\Xi = CP^{-1} \sum_{k=0}^{M-1} (R^T)^k A$ then the system doesn't produce any overload events.

*Proof.* Let $\bar{x}(t)$ be the users' rates vector. In order to serve this data flow the system allocates the vector of service rates $\bar{u}(t)$ such that $Ax(t) - (I - R^T)P\bar{u}(t) = 0$. It is always possible if $\bar{u}(t) = P^{-1}(I - R^T)^{-1}A\bar{x} \in U$. The inverse matrix exists as a power series $[I - R]^{-1} = \sum_{k=0}^{M-1} R^k$. Stability condition is a formal expression of following inclusion: $P^{-1}(I - R^T)^{-1}A\bar{x} \in \text{int } U$, where boundary of $U$ was excluded to prevent overload event. Matrix $\Xi$ describes controllability of the system, defined by matrixes $C, P, R, A$.

To clarify this notation and stability condition we will now consider some classical examples.

## 2.2 Examples

**Single server.** This is the simplest possible network. This model fits well for investigation interaction between users rather then network dynamic. Consider single node with capacity $p$, $M = L = 1$. There are $N$ users with rates $\bar{x}(t) = (x_1(t), ..., x_N(t))^T$. Matrix $A = \begin{bmatrix} 1 & ... & 1 \end{bmatrix}$, and set $U = [0, p] \subset R_+$. Here $\Xi = \begin{bmatrix} \frac{1}{p} \end{bmatrix}$ and stability condition is $x_1(t) + ... + x_N(t) < p$.

**Klimov model.** Consider a model shown in Fig. 1. There are $N$ users with vector of rates $\bar{x}(t) = (x_1(t), ..., x_N(t))^T$ and $N$ links, where their data flows are processed. Each link has maximum capacity $p_j$, $j = 1, .., N$ and $u_j(t)$ is percentage of use of total amount of resource (e.g. CPU time, network bandwidth). Taking into account that $C = \begin{bmatrix} 1 & ... & 1 \end{bmatrix}$, $P = \begin{bmatrix} p_1 & ... & 0 \\ \vdots & \ddots & \vdots \\ 0 & ... & p_N \end{bmatrix}$ we have

$$U = \left\{ u \in R_+^K \mid CP^{-1}u \leq 1 \right\} = \left\{ u \in R_+^K \mid \frac{u_1}{p_1} + ... + \frac{u_N}{p_N} \leq 1 \right\}.$$

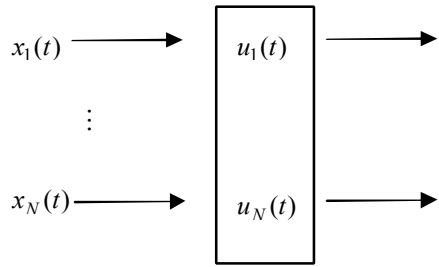

Fig. 1. Klimov model

Matrix $A = I$, so $\Xi = \begin{bmatrix} \frac{1}{p_1} & \ldots & \frac{1}{p_N} \end{bmatrix}$ and stability condition is given by the following formula $\frac{x_1(t)}{p_1} + \ldots + \frac{x_N(t)}{p_N} < 1$

**Simple re-entrant line.** Consider a more complex system shown on Fig. 2.

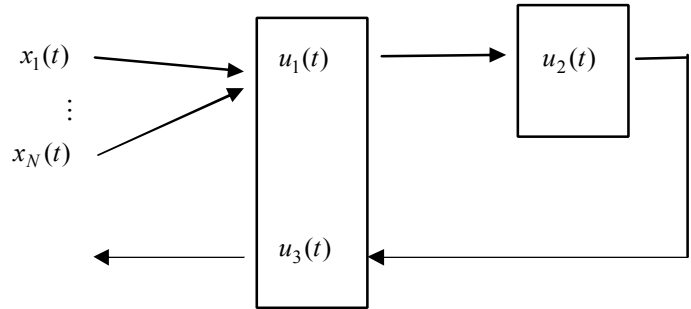

Fig. 2. Re-entrant line.

There are $N$ users with vector of rates $\bar{x}(t) = (x_1(t),\ldots,x_N(t))^T$ sending data to first link of first node with $p_1$ maximum capacity and $u_1(t) \in [0,1]$ control parameter. After that packets go to second node with the single link with $p_2$ maximum capacity and $u_2(t) \in [0,1]$ control parameter. Finally, packets return to second link of first node ($p_3$ maximum capacity and $u_3(t) \in [0,1]$ control parameter) and left the system. Links 1 and 2 are situated on the same node, so their summary capacity is limited.

Let us define all matrixes for this network:
$$P = \begin{bmatrix} p_1 & 0 & 0 \\ 0 & p_2 & 0 \\ 0 & 0 & p_3 \end{bmatrix}, C = \begin{bmatrix} 1 & 0 & 1 \\ 0 & 1 & 0 \end{bmatrix}, A = \begin{bmatrix} 1 & \ldots & 1 \\ 0 & \ldots & 0 \\ 0 & \ldots & 0 \end{bmatrix},$$
$$U = \left\{ u \in R_+^3 \mid \frac{u_1}{p_1} + \frac{u_3}{p_3} \leq 1, u_2 \leq p_2 \right\}$$

Routing matrix $R^T = \begin{bmatrix} 0 & 0 & 0 \\ 1 & 0 & 0 \\ 0 & 1 & 0 \end{bmatrix}$, and $(I - R^T)^{-1} = I + R^T = \begin{bmatrix} 1 & 0 & 0 \\ 1 & 1 & 0 \\ 1 & 1 & 1 \end{bmatrix}$.

Finally, $\Xi = \begin{bmatrix} \frac{1}{p_1}+\frac{1}{p_3} & \cdots & \frac{1}{p_1}+\frac{1}{p_3} \\ \frac{1}{p_2} & \cdots & \frac{1}{p_2} \end{bmatrix}$ and stability condition consists of two inequalities:

$$\left(\frac{1}{p_1}+\frac{1}{p_3}\right)(x_1(t)+\ldots x_N(t)) < 1$$

$$(x_1(t)+\ldots x_N(t)) < p_2$$

If rates $x$ satisfy stability condition then network will be lossless. But from practical point of view, there are many problems with applicability of this condition. First, in real network each user doesn't have information about system's current state and about rates of other users so he cannot calculate a proper rate. Second, the user cannot choose any rate he wants (at least in TCP scheme). Instead he chooses protocol, controlling his rate. The dynamic of protocol is investigated in next section.

## 2.3 System dynamic

There are different instances of TCP class. The mostly used one is New-Reno. The behavior of New Reno is close to pure AIMD scheme. It adapts to the available capacity by increasing the window size in a linear way by $\alpha$ packets every round trip time and when it detects congestion it decreases the window size to $\beta$ times its value. The constants $\alpha$ and $\beta$ are 1 and 1/2, respectively, in New-Reno.

The original TCP specification doesn't forbid using any user defined congestion control mechanism. Even for AIMD like control one have freedom in changing values $\alpha$ and $\beta$. Obviously, if you set $\alpha$ and $\beta$ bigger than standard (1, 1/2), you receive advantage against flow with lower values. This will cause unfair allocation of network resources and thus is undesirable.

This interaction of first protocol on second one is called unfriendly and it is said that the first protocol is more "aggressive" and latter is more "peaceful". Aggression here is an opportunity to grab more network resourses than would be fair. The question of protocols interaction is quite complex. Building analytic model for predicting network behavior for different protocols is a challenging problem.

In last years, more aggressive TCP versions have appeared, such as HSTCP (High Speed TCP) and Scalable TCP. HSTCP can be modeled by an AIMD behavior where $\alpha$ and $\beta$ are not constant anymore: $\alpha$ and $\beta$ have minimum values of 1 and of 1/2, resp. and both increase as the window size increases. Scalable TCP is an MIMD (Multiplicative Increase Multiplicative Decrease) protocol, where the window size increases exponentially instead of linearly and is thus more aggressive. Versions of TCP which are less aggressive than the New-Reno also exist, such as Vegas.

In this section we build a dynamical model of AIMD connection using differential equation with discontinuous right-hand side.

Let $\bar{x}_0$ be an initial vector of rates and $\bar{\alpha}$, $\bar{\beta}$ vectors of parameters. According to original AIMD scheme [1] user rates are increasing between overloads with rate $\bar{\alpha}$. When overload occurs the rate drops to $\bar{\beta}x$. Now we will put into formal definitions. Define $t_i$, $i \geq 1$ as the first moment of time $t_i > t_{i-1}$, such that there exists $j \in J$: $[\Xi x(t_i)]_j = 1$. We will assume that the RTT (round trip times) are the same for all connections and losses are synchronized: when the combined rates attain capacity, all connections suffer from a loss.

Consider the following equation

$$\dot{x}(t) = \bar{\alpha} - \sum_{i=1}^{N_t}(I-B)x(t_i)\delta(t-t_i), \tag{1}$$

where $\delta$ is delta-function, $B = diag\{\beta_1,...,\beta_N\}$, $N_t = \max\{n : t_n \leq t\}$. Equation (1) is well-defined Caratheodory equation with discontinuous right-hand side, differential equations with impulses have been examined in many papers, which cannot all be referenced here (see [11] for references). It is known [11] that there is an almost continuous solution (continuous in all points except a set of measure zero)

$$x(t) = \alpha t - \sum_{i=1}^{N_t}(I-B)x(t_i)\eta(t-t_i), \tag{2}$$

where $\eta$ is the Heaviside step function. Explicit formula (2) is not very practical but gives us important information about solution existence and its continuity in almost all points.

Denote $V$ as a set $\{v \in R_+^N \mid \min[\Xi v]_j = 1, j \in J\}$ and $W$ as a set $\{w \in R_+^N \mid \Xi w \leq 1\}$. It is clear that $V$ is compact set, $W$ is convex compact set and $V \subseteq \partial W$.

**Condition 2.1.** For any $x \in V$ it is true that $Bx \in \text{int}\,W$.

Let us explain Condition 2.1 informally. $W$ is the vector set of possible user rates. $W$ is convex compact set and $x(t) \in W$ for $t \geq t_0$. As mentioned $x(t)$ is an almost continuous function, and drops happened when $x(t) \in V$. After drop event users rates equal to $Bx(t)$. The condition 2.1 means that after applying decreasing operator $B$ user rate still will be in the admissible set $W$.

We use Brouwer's theorem to prove existence of limit solution of (1). Let us remind one version of it [12]. First we define homeomorphism and simplexes.

A set $X$ is homeomorphic to the set $Y$ if there is a bijective continuous function $h : X \to Y$ such that $h^{-1}$ is also continuous.

A set $\{x^0,...,x^n\} \subset R^m$ is affinely independent if $\sum_{i=0}^{n} \lambda_i x^i = 0$ and $\sum_{i=0}^{n} \lambda_i = 0$ imply that $\lambda_0 = ... = \lambda_n = 0$. An $n$-simplex is the set of all positive convex combinations of $n+1$ element affinely independent set. Let $\Delta_n$ denote the closure of the standard $n$-simplex $\left\{ y \in R^{n+1} \mid y_i > 0, i = 0,...,n; \sum_{i=0}^{n} y_i = 1 \right\}$.

**Theorem** (Brouwer) [12] Let $X \subset R^n$ be homeomorphic to simplex $\Delta_{n-1}$ and let $f: X \to X$ be continuous, then $f$ has a fixed point.

### 2.4  Solution of dynamic system

Now we can formulate the main result of this section – existence and uniqueness of the limit solution.

**Proposition 2.2.** Let us consider admissible pair $\bar{\alpha}$, $\bar{\beta}$. If Condition 2.1 holds then for any $\bar{x}_0 \in W$ solution of (1) exists and is converging to unique periodical solution $\hat{x}(t)$.

*Proof.* Consider a map $f: V \to V$ defined as $f(v) = \{y \in V \mid \exists t > 0 : Bv + \alpha t = y\}$. Condition 2.1 holds, this means that $f(\cdot)$ is the well-defined function. It is clear that $f(\cdot)$ is continuous. Consider simplex $\Delta_{N-1}$ defined as $co\{e_1,...,e_N\}$, where $e_i$ are vectors from standard basis. For any $x \in V$ there exists unique $\psi \in \Delta_{N-1}$ such that $x = a\psi$, for some $a \in R$. This means that $V$ is homeomorphic to $\Delta_{N-1}$. Then Brouwer's theorem is applicable and there is a fixed point. Let $\bar{x}^* \in V$ be the fixed point of the map $f(\cdot)$. Denote $\bar{\gamma}$ as a vector with components $\gamma_i = \dfrac{\alpha_i}{1-\beta_i}$. Then the following is true: $B\bar{x}^* + \bar{\alpha}T = \bar{x}^*$, where $T = \min\{t : \bar{x}^* + \bar{\alpha}t \in V\}$. Using this property, we can calculate $\bar{x}^*$ directly

$$\bar{x}^* = (I-B)^{-1}\bar{\alpha}T = \gamma T.$$

Condition $\bar{x}^* + \bar{\alpha}T \in V$ could be re-written as $\min_i (\Xi \bar{x}^*)_i = 1$ or $\min_i (\Xi \gamma T)_i = 1$. So we can conclude that $T$ is unique, so fixed point also is unique. Now let us solve (1) with the initial point $B\bar{x}^*$. It is clear that $\hat{x}(t) = \alpha t + B\bar{x}^*$ for $t \in [0,T)$ and $\hat{x}(T) = \hat{x}(0)$. So, solution $\hat{x}(t)$ is periodical with period $T$. Consider arbitrary solution with the initial point $x(0) \in W$. Let $x(t_1)$ be the first moment when $x(t_1) \in V$, then define $z_n = f(z_{n-1})$, $z_0 = f(x(t_1))$. All elements of $\{z_n\}$, $n = 0,...,\infty$ belong to compact set $V$ so there is a limit $\tilde{x} \in V$, $\tilde{x} = f(\tilde{x})$. The only possible solution, asso-ciated with this limit point is $\hat{x}(t)$. So we proved convergence of any solution to this periodical solution.

In next section we solve dynamic system (1) for different examples using Proposition 2.2. This notation is as follows: $T$ - oscillation period, $x^*$ fixed point from set $V$.

## 2.5 Applications

**Single server.** Note that $\Xi = \begin{bmatrix} \frac{1}{p} & \ldots & \frac{1}{p} \end{bmatrix}$, so condition $x \in V$ could be re-written as $\frac{x_1}{p} + \ldots + \frac{x_N}{p} = 1$. Let $x^* \in V$ be fixed point, then $x^* = \begin{bmatrix} \frac{\alpha_1}{1-\beta_1} & \ldots & \frac{\alpha_N}{1-\beta_N} \end{bmatrix}^T T$, and $\Xi A(Bx^* + \alpha T) = 1$. Solving this system of equations we obtain:

$$T = \frac{p}{\frac{\alpha_1}{1-\beta_1} + \ldots + \frac{\alpha_N}{1-\beta_N}}.$$

Note, if all $\alpha_i = \alpha$, $\beta_i = \beta$, then $T = \frac{p(1-\beta)}{N\alpha}$, $x_i^* = \frac{p}{N}$.

This result was found by solving difference equation directly for two protocols system considered in [Altman].

**Klimov model.** Consider a model shown in Fig. 1. Matrix $\Xi = \begin{bmatrix} \frac{1}{p_1} & \ldots & \frac{1}{p_N} \end{bmatrix}$, so condition $x \in V$ could be re-written as $\frac{x_1}{p_1} + \ldots + \frac{x_N}{p_N} = 1$. Fixed point is the solution of following system of equations:

$$\Xi(Bx + \alpha T) = \frac{\beta_1 x_1}{p_1} + \frac{\alpha_1}{p_1} T + \ldots \frac{\beta_N x_N}{p_N} + \frac{\alpha_N}{p_N} T = 1,$$

$$x = \begin{bmatrix} \frac{\alpha_1}{1-\beta_1} & \ldots & \frac{\alpha_N}{1-\beta_N} \end{bmatrix}^T T.$$

Finally,

$$T = \frac{1}{\frac{\alpha_1}{(1-\beta_1)p_1} + \ldots + \frac{\alpha_N}{(1-\beta_N)p_N}}, \quad x_i^* = \frac{\alpha_i}{(1-\beta_i)\left(\frac{\alpha_1}{(1-\beta_1)p_1} + \ldots + \frac{\alpha_N}{(1-\beta_N)p_N}\right)}.$$

**Re-entrant line.**

$$\Xi = \begin{bmatrix} \frac{1}{p_1}+\frac{1}{p_3} & \cdots & \frac{1}{p_1}+\frac{1}{p_3} \\ \frac{1}{p_2} & \cdots & \frac{1}{p_2} \end{bmatrix}.$$ Fixed point is the solution of following system of equations:

$$\Xi(Bx+\alpha T) = \begin{bmatrix} \left(\frac{1}{p_1}+\frac{1}{p_3}\right)(\beta_1 x_1+\alpha_1 T+\ldots+\beta_N x_N+\alpha_N T) \\ \frac{1}{p_2}(\beta_1 x_1+\alpha_1 T+\ldots+\beta_N x_N+\alpha_N T) \end{bmatrix} = 1,$$

$$x = \begin{bmatrix} \frac{\alpha_1}{1-\beta_1} & \cdots & \frac{\alpha_N}{1-\beta_N} \end{bmatrix}^T T$$

The result is

$$T = \frac{1}{\sum_i \frac{\alpha_i}{(1-\beta_i)} \max\left\{\frac{1}{p_2}, \frac{1}{p_1}+\frac{1}{p_3}\right\}}, \quad x_i^* = \frac{\alpha_i}{(1-\beta_i)} T.$$

## 3  AIMD game formulation

Now we consider a competition between users which use AIMD version of TCP with different parameters. Their connections are sharing a common network. We will assume that users send their packets exactly the same way, so we can reduce network topology to the single link type with capacity $c$.

In order to formulate a game in strategic form we must specify the players, their strategies, and their potential payoffs. Player here is a user. We assume that there are $N$ AIMD strategies $s_i$ with control parameters $(\alpha_i, \beta_i)$, $i=1,\ldots,N$. Denote $S$ as a set of all possible strategies.

We consider payoff of the form $J_i(s) = Thp_i(s) - \lambda R(s)$, where $\bar{s} = (s_1,\ldots,s_N)$ - vector of strategies; $Thp_i(s) = 0.5(1+\beta_i)x_i^*$ - average throughput of $i$'s player; $\lambda$ - tradeoff parameter (sensitivity to losses); $R(s) = \frac{1}{T(s)}$ - loss rate.

**Example.** Let us calculate payoffs for two strategies:

$$J_1(s_i, s_i) = J_2(s_i, s_i) = \frac{(1+\beta_i)}{4}c - \lambda \frac{2\alpha_i}{c\bar{\beta}_i},$$

$$J_1(s_1, s_2) = \frac{(1+\beta_1)\alpha_1 c \bar{\beta}_2}{2(\alpha_1 \bar{\beta}_2 + \alpha_2 \bar{\beta}_1)} - \frac{\lambda}{c}\left(\frac{\alpha_1}{\bar{\beta}_1} + \frac{\alpha_2}{\bar{\beta}_2}\right),$$

$$J_1(s_2, s_1) = \frac{(1+\beta_2)\alpha_2 c \bar{\beta}_1}{2(\alpha_1 \bar{\beta}_2 + \alpha_2 \bar{\beta}_1)} - \frac{\lambda}{c}\left(\frac{\alpha_1}{\bar{\beta}_1} + \frac{\alpha_2}{\bar{\beta}_2}\right),$$

$$J_2(s_1, s_2) = J_1(s_2, s_1),$$
$$J_2(s_2, s_1) = J_1(s_1, s_2)$$

### 3.1 Equilibrium in $N$ protocols game.

Consider a game with $N$ AIMD strategies. We assume that all $s_i$ are ordered lexicographically, $s_1 \geq s_2 \geq ... \geq s_N$, where $s_i \geq s_j$ means that $\alpha_i \geq \alpha_j$ and $\beta_i \geq \beta_j$. In other words protocols are sorted by aggressiveness ordering.

**Proposition 3.1.** If $\lambda$ is sufficiently small than the most aggressive protocol is dominant strategy.

*Proof.* Suppose $\alpha_1 \geq \alpha_i$, $\beta_1 \geq \beta_i$ for all $i = 2,..., N$. Consider payoffs for the first player $J_1(s_1, s_{-1})$ and $J_1(s_j, s_{-1})$. Let us find the period for both strategy profiles:

$$T(s_1, s_{-1}) = \frac{c}{\frac{\alpha_1}{(1-\beta_1)} + A}, \text{ where } A = \sum_k \frac{\alpha_k}{1-\beta_k} \text{ is sum, defined by strategy set}$$

$s_{-1}$, $T(s_j, s_{-1}) = \dfrac{c}{\dfrac{\alpha_j}{(1-\beta_j)} + A}$. Note, that $T(s_1, s_{-1}) < T(s_j, s_{-1})$.

Calculate throughputs:

$$Thp_1(s_1, s_{-1}) = \frac{(1+\beta_1)x_1^*}{2} = \frac{(1+\beta_1)\alpha_1 c}{2(1-\beta_1)\left(\frac{\alpha_1}{(1-\beta_1)} + A\right)} = \frac{(1+\beta_1)c}{2\left(1 + \frac{(1-\beta_1)}{\alpha_1}A\right)},$$

$$Thp_1(s_j, s_{-1}) = \frac{(1+\beta_j)x_j^*}{2} = \frac{(1+\beta_j)c}{2\left(1 + \frac{(1-\beta_j)}{\alpha_j}A\right)}.$$

Calculate payoffs:

$$J_1(s_1, s_{-1}) = Thp_1(s_1, s_{-1}) - \frac{\lambda}{c}\left(\frac{\alpha_1}{(1-\beta_1)} + A\right),$$

$$J_1(s_j, s_{-1}) = Thp_1(s_j, s_{-1}) - \frac{\lambda}{c}\left(\frac{\alpha_j}{(1-\beta_j)} + A\right).$$

$$Thp_1(s_1, s_{-1}) - \frac{\lambda}{c}\left(\frac{\alpha_1}{(1-\beta_1)} + A\right) > Thp_1(s_j, s_{-1}) - \frac{\lambda}{c}\left(\frac{\alpha_j}{(1-\beta_j)} + A\right),$$

$$\frac{(1+\beta_1)c}{2\left(1 + \frac{(1-\beta_1)}{\alpha_1}A\right)} - \frac{(1+\beta_j)c}{2\left(1 + \frac{(1-\beta_j)}{\alpha_j}A\right)} > \frac{\lambda}{c}\frac{\alpha_1}{(1-\beta_1)},$$

$$\lambda < \frac{c^2(1-\beta_1)}{\alpha_1}\left[\frac{(1+\beta_1)c}{2\left(1+\frac{(1-\beta_1)}{\alpha_1}A\right)} - \frac{(1+\beta_j)c}{2\left(1+\frac{(1-\beta_j)}{\alpha_j}A\right)}\right]$$

And since expression in right side is positive we obtain the result.

### 3.2 Nash mixed and pure in two protocols game.

Here we investigate the game for two protocols and find conditions for Nash equilibrium. From definition it is clear that $J_i(s_k, s_k) = J_j(s_k, s_k)$ - we will write just $J(s_k, s_k)$, $J_i(s_k, s_p) = J_j(s_p, s_k), j \in \{1,2\} \setminus i$.

The matrix of this game is shown in Table.

Table 1.

Player 2

| | Strategy | $s_1$ | $s_2$ |
|---|---|---|---|
| Player 1 | $s_1$ | $J(s_1,s_1)$, $J(s_1,s_1)$ | $J(s_1,s_2)$, $J(s_2,s_1)$ |
| | $s_2$ | $J(s_2,s_1)$, $J(s_1,s_2)$ | $J(s_2,s_2)$, $J(s_2,s_2)$ |

Using standard techniques for calculating Nash we obtain:
$J_1(s_1) = pJ(s_1,s_1) + (1-p)J(s_1,s_2)$
$J_1(s_2) = pJ(s_2,s_1) + (1-p)J(s_2,s_2)$

assuming the probability of player 2 using the first strategy is $p$. In Nash equilibrium the payoff can't be further increased, so these two values should be indistinguishable, which leads to the following equation

$$pJ(s_1,s_1) + (1-p)J(s_1,s_2) = pJ(s_2,s_1) + (1-p)J(s_2,s_2)$$

Or, after solving it for $p$:

$$p = \frac{J(s_1,s_2) - J(s_2,s_2)}{(J(s_1,s_2) - J(s_2,s_2)) + (J(s_2,s_1) - J(s_1,s_1))}$$

Taking into account that $p$ is a probability, we impose a natural restrictions on it: $0 \le p \le 1$, where cases with $p = 1$ or $p = 0$ result in game having a pure-strategy equilibrium (with dominant strategy $s_1$ and $s_2$, respectively), and $0 < p < 1$ corresponds to the case of mixed-strategy Nash equilibrium.

Should we investigate the conditions for the former, we get

$$J(s_1,s_2) - J(s_2,s_2) = \frac{-\lambda(\alpha_2(1-\beta_1)+\alpha_1(1-\beta_2))}{c(1-\beta_2)(1-\beta_1)} + \frac{c\alpha_1(1+\beta_1)(1-\beta_2)}{2(\alpha_2(1-\beta_1)+\alpha_1(1-\beta_2))} +$$

$$\frac{2\lambda\alpha_2}{c(1-\beta_2)} + \frac{1}{4}C(1+\beta_2).$$

$$J(s_2,s_1) - J(s_1,s_1) = \frac{-\lambda(\alpha_2(1-\beta_1)+\alpha_1(1-\beta_2))}{c(1-\beta_2)(1-\beta_1)} + \frac{c\alpha_2(1+\beta_2)(1-\beta_1)}{2(\alpha_2(1-\beta_1)+\alpha_1(1-\beta_2))} +$$

$$\frac{2\lambda\alpha_1}{c(1-\beta_1)} + \frac{1}{4}C(1+\beta_1)$$

Consequently,

$$p = 1 - \frac{2\alpha_2(1-\beta_1)}{\alpha_2(1-\beta_1)-\alpha_1(1-\beta_2)} - \frac{4\lambda(\alpha_1+\alpha_2)+c^2(\beta_1^2-1)}{c^2(1-\beta_1)(\beta_1-\beta_2)} + \frac{4\lambda\alpha_2}{c^2(1-\beta_1)(1-\beta_2)}.$$

Considering the case where game has pure-strategy equilibrium, we get two possible conditions: $p=1$ or $p=0$.

Solving the equations, we find the values of $\lambda$ that correspond to the case of dominant strategy:

$$\lambda = \frac{c\beta_1\beta_2(\alpha_2\beta_1(1-\beta_1+2\beta_2)-\alpha_1(1-\beta_2)(1+\beta_1))}{4(\alpha_2(1-\beta_1)-\alpha_1(1-\beta_2))(\alpha_2(1-\beta_1)+\alpha_1(1-\beta_2))}$$

$$\lambda = \frac{c\beta_1\beta_2(\alpha_2\beta_1(1+\beta_2)+\alpha_1(1-\beta_2)(1+2\beta_1-\beta_2))}{4(\alpha_2(1-\beta_1)-\alpha_1(1-\beta_2))(\alpha_2(1-\beta_1)+\alpha_1(1-\beta_2))} \quad (3)$$

Now, for the game to have mixed-strategy equilibrium the following system of inequalities must hold:

$p < 1$ and $p > 0$

After solving this system for $\lambda$ we get

$$\frac{C^2\overline{\beta}_1\overline{\beta}_2(\alpha_1\overline{\beta}_2(1+\beta_1)+\alpha_2\overline{\beta}_1(\beta_1-2\beta_2-1))}{4(\alpha_1^2\overline{\beta}_2^2-\alpha_2^2\overline{\beta}_1^2)} < \lambda < \frac{C^2\overline{\beta}_1\overline{\beta}_2(\alpha_1\overline{\beta}_2(1+2\beta_1-\beta_2)-\alpha_2\overline{\beta}_1(1+\beta_2))}{4(\alpha_1^2\overline{\beta}_2^2-\alpha_2^2\overline{\beta}_1^2)}$$

(4)

We have proved:

**Proposition 3.2.** If $\lambda$ satisfies (4) then there is Nash equilibrium in mixed strategies. If $\lambda$ satisfies (3) then there is Nash equilibrium in pure strategies

### 3.3 Extension for protocols parameters.

The game settings in previous sections were limited by aggressive ordering of protocols. In this section we weaken this condition to cover protocol parameters relation that falls beyond the "aggressive-peaceful" scheme, namely situation when $\alpha_1 \leq \alpha_2$ and $\beta_1 \geq \beta_2$.

Applying the same considerations as above, we get the same results for pure-strategy Nash equilibria, but for mixed-strategy equilibrium an additional constraint emerges.

Since we're looking for cases with $0 < p < 1$, we get the following conditions for $p > 0$:

$$\begin{cases} J(s_1, s_2) - J(s_2, s_2) > 0 \\ (J(s_1, s_2) - J(s_2, s_2)) + (J(s_2, s_1) - J(s_1, s_1)) > 0 \end{cases}$$

Similarly, $p < 1$ holds when

$$(J(s_1, s_2) - J(s_2, s_2)) + (J(s_2, s_1) - J(s_1, s_1)) > J(s_1, s_2) - J(s_2, s_2)$$

(It can be shown that other case with $J(s_1, s_2) - J(s_2, s_2) < 0$ results $\lambda < 0$, which has no physical sense, recalling that $\lambda$ is an error weight).

So, in the end we have the following system of inequalities:

$$\begin{cases} J(s_1, s_2) - J(s_2, s_2) > 0 \\ J(s_2, s_1) - J(s_1, s_1) > 0 \end{cases}$$

Or, after replacement of $J$ and transformations

$$\begin{cases} \lambda < \dfrac{C^2 \bar{\beta}_1 \bar{\beta}_2 (\alpha_1 \bar{\beta}_2 (1 + 2\beta_1 - \beta_2) - \alpha_2 \bar{\beta}_1 (1 + \beta_2))}{4(\alpha_1^2 \bar{\beta}_2^2 - \alpha_2^2 \bar{\beta}_1^2)} \\ \lambda > \dfrac{C^2 \bar{\beta}_1 \bar{\beta}_2 (\alpha_1 \bar{\beta}_2 (1 + \beta_1) + \alpha_2 \bar{\beta}_1 (\beta_1 - 2\beta_2 - 1))}{4(\alpha_1^2 \bar{\beta}_2^2 - \alpha_2^2 \bar{\beta}_1^2)} \end{cases}$$

Replacing

$$\mu_1 = \frac{C^2 \bar{\beta}_1 \bar{\beta}_2 (\alpha_1 \bar{\beta}_2 (1 + 2\beta_1 - \beta_2) - \alpha_2 \bar{\beta}_1 (1 + \beta_2))}{4(\alpha_1^2 \bar{\beta}_2^2 - \alpha_2^2 \bar{\beta}_1^2)}$$

$$\mu_2 = \frac{C^2 \bar{\beta}_1 \bar{\beta}_2 (\alpha_1 \bar{\beta}_2 (1 + \beta_1) + \alpha_2 \bar{\beta}_1 (\beta_1 - 2\beta_2 - 1))}{4(\alpha_1^2 \bar{\beta}_2^2 - \alpha_2^2 \bar{\beta}_1^2)}$$

We get two possible solutions to the system above:

$$\begin{cases} \alpha_2(1 - \beta_1) - \alpha_1(1 - \beta_2) > 0 \\ \mu_2 < \lambda < \mu_1 \end{cases}$$

Or

$$\begin{cases} \alpha_2(1 - \beta_1) - \alpha_1(1 - \beta_2) < 0 \\ \mu_1 < \lambda < \mu_2 \end{cases}$$

Since $\alpha_1 \leq \alpha_2$ and $\beta_1 \geq \beta_2$ then $\mu_2 > \mu_1$, the actual solution is

$$\begin{cases} \alpha_2(1 - \beta_1) - \alpha_1(1 - \beta_2) < 0 \\ \mu_1 < \lambda < \mu_2 \end{cases}$$

**Proposition 3.3.** If $\alpha_1 \leq \alpha_2$ and $\beta_1 \geq \beta_2$ and $\alpha_1 < \alpha_2 < \dfrac{\alpha_1(1-\beta_2)}{1-\beta_1}$, $\mu_1 < \lambda < \mu_2$,

then there is evolutionary stable equilibrium in mixed strategies.

Formulated conditions are consistent with the previous result with regards to protocol parameters specifics.

In next section we illustrate theoretical results with numeric simulations.

## 4 Simulation

We study in this section numerically dynamic system (1) and equilibriums of defined game with replicator dynamics. The practical value of these results could be divided on two parts. Firstly, this is analytical tool for predicting shares of network resources for given set of AIMD protocols. Existence and uniqueness of this point of resource allocation proved in Proposition 2.2.

Secondly, we can model users behavior (taking into account usual game theory assumption about rationality, common knowledge etc.) using replicator dynamic equation. This equation is rather quality solution tool that show a dynamic and shares of network resources for each users group.

### 4.1 Solution for dynamic system

Numerical simulations were made using Wolfram Mathematica environment. On the picture below we show convergence of AIMD scheme for 2 and 3 dimensions.

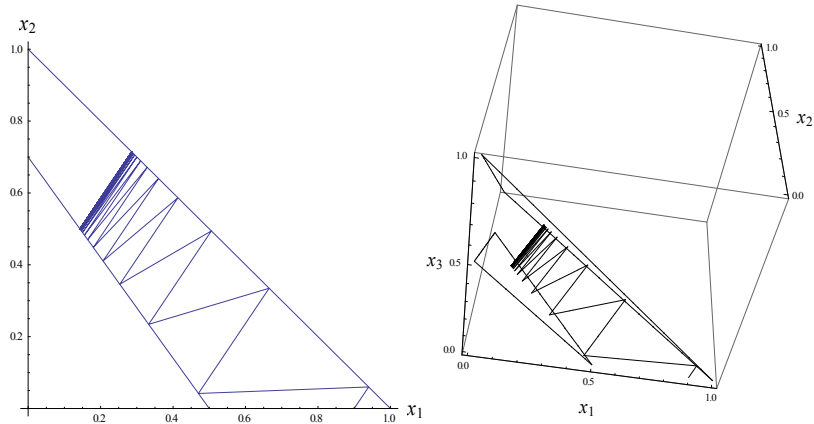

Fig. 3. Simulations results for 2-d and 3-d systems

### 4.2 Replicator dynamic

We introduce here the replicator dynamics which describes the evolution in the population of the various strategies. In the replicator dynamics, the share of a strategy in the population grows at a rate equal to the difference between the payoff of that strategy and the average payoff of the population. More precisely, consider N strategies. Let $x$ be the N-dimensional vector whose $i$ element $x_i$ is the population share of strategy i (i.e. the fraction of the population that uses strategy $i$). Thus, we have $\sum_i x_i(t) = 1$, $x_i(t) \geq 0$. Then the replicator dynamics is defined as

$$\dot{x}_i(t) = x_i(t) K \left( \sum_{j \neq i} J(i,j) x_j - \sum_j x_j(t) \sum_{k \neq j} J(j,k) x_k \right)$$

We investigate a case with $N = 3$ distinct strategies and pairwise payoff comparison:

$J(s_1, X(t-\tau)) = x_1(t-\tau) J(s_1, s_1) + x_2(t-\tau) J(s_1, s_2) + x_3(t-\tau) J(s_1, s_3)$

$\dot{x}_i(t) = x_i(t) K (J(i, X(t-\tau) - (x_1(t-\tau) J(s_1, X(t-\tau)) + x_2(t-\tau) J(s_1, X(t-\tau)) +$
$+ x_2(t-\tau) J(s_1, X(t-\tau))))$

We provide a simulation results for 3 sets of parameters:

| $\alpha_1$ | $\alpha_2$ | $\alpha_3$ | $\beta_1$ | $\beta_2$ | $\beta_3$ | $\lambda$ | $K$ | $c$ | $\tau$ |
|---|---|---|---|---|---|---|---|---|---|
| 1.5 | 1.25 | 1 | 0.75 | 0.5 | 0.25 | 168 | 0.2 | 50 | 0.25 |

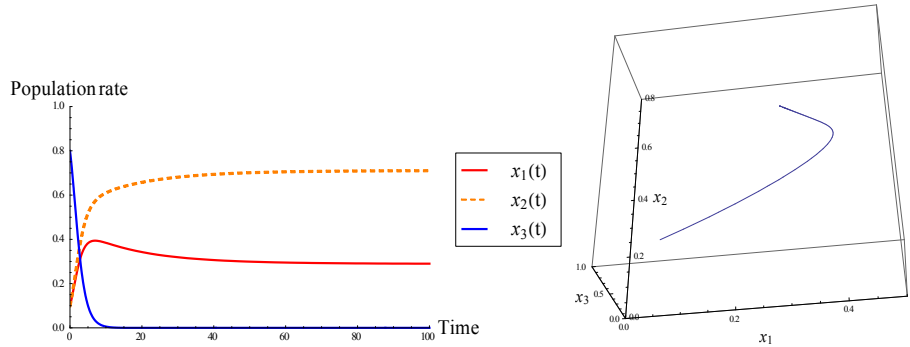

Fig. 4 $x_1(t)$, $x_2(t)$, $x_3(t)$ - shares of population

| $\alpha_1$ | $\alpha_2$ | $\alpha_3$ | $\beta_1$ | $\beta_2$ | $\beta_3$ | $\lambda$ | $K$ | $c$ | $\tau$ |
|---|---|---|---|---|---|---|---|---|---|
| 1.5 | 1.25 | 1 | 0.75 | 0.5 | 0.25 | 140 | 0.2 | 50 | 0.25 |

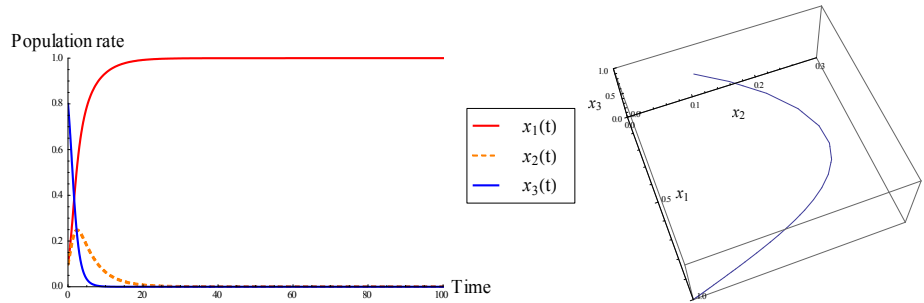

Fig. 5 $x_1(t)$, $x_2(t)$, $x_3(t)$

| $\alpha_1$ | $\alpha_2$ | $\alpha_3$ | $\beta_1$ | $\beta_2$ | $\beta_3$ | $\lambda$ | $K$ | $c$ | $\tau$ |
|---|---|---|---|---|---|---|---|---|---|
| 1.5 | 1.25 | 1 | 0.75 | 0.5 | 0.25 | 168 | 0.2 | 50 | 15 |

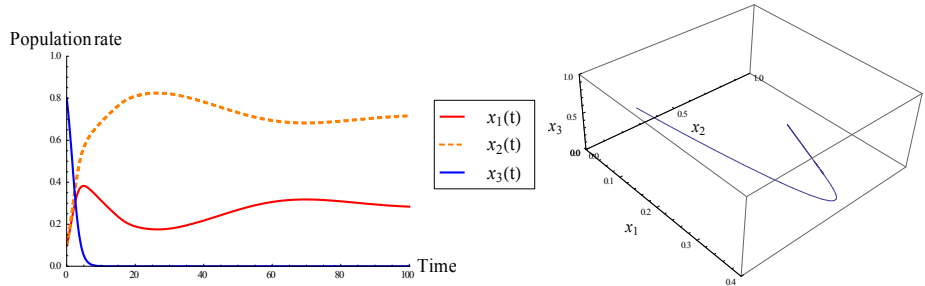

Fig. 6 $x_1(t)$, $x_2(t)$, $x_3(t)$

Expand the replicator dynamics differential equation to include a case with $N = 3$ different strategies.

$$J(s_1, X(t-\tau)) = \sum_p \sum_q x_p(t-\tau) x_q(t-\tau) J(A, p, q)$$

We provide a simulation results for 2 sets of parameters:

| $\alpha_1$ | $\alpha_2$ | $\alpha_3$ | $\beta_1$ | $\beta_2$ | $\beta_3$ | $\lambda$ | $K$ | $c$ | $\tau$ |
|---|---|---|---|---|---|---|---|---|---|
| 1.3 | 1.5 | 1.1 | 0.25 | 0.4 | 0.85 | 140 | 0.25 | 50 | 5 |

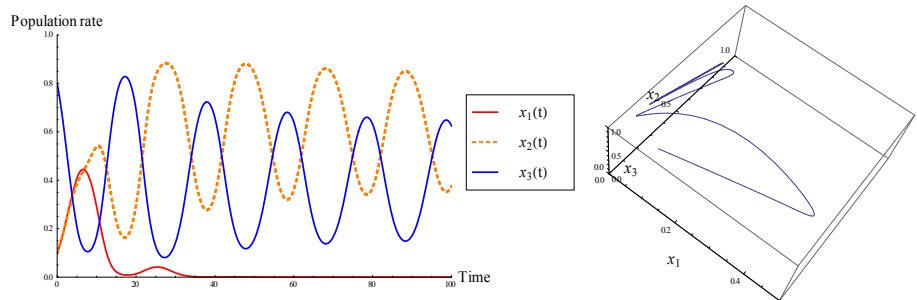

Fig. 7 $x_1(t)$, $x_2(t)$, $x_3(t)$

| $\alpha_1$ | $\alpha_2$ | $\alpha_3$ | $\beta_1$ | $\beta_2$ | $\beta_3$ | $\lambda$ | $K$ | $c$ | $\tau$ |
|---|---|---|---|---|---|---|---|---|---|
| 1.5 | 1.25 | 1 | 0.75 | 0.5 | 0.25 | 140 | 0.25 | 50 | 5 |

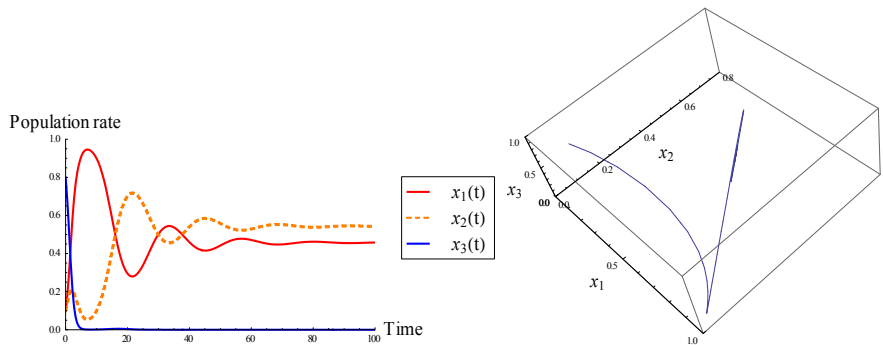

Fig. 8 $x_1(t)$, $x_2(t)$, $x_3(t)$

## 5 Conclusions

This paper deals with modeling of network's dynamic using evolutionary games approach. It is presented the model of network's dynamic using differential equations with discontinuous right side and proved existence and uniqueness of solution, formulated payoff matrix for a network AIMD connections game and found conditions of equilibrium existence depending of loss sensitivity parameter. The results are illustrated by simulations using Wolfram Mathematica.

## 6 References

1. Chiu, Dah-Ming, and Raj Jain. Analysis of the increase and decrease algorithms for congestion avoidance in computer networks. Computer Networks and ISDN systems 17.1 (1989): 1-14.
2. Kelly, Frank P., Aman K. Maulloo, and David KH Tan. "Rate control for communication networks: shadow prices, proportional fairness and stability." Journal of the Operational Research society (1998): 237-252.
3. Mo J., Walrand J. Fair end-to-end window-based congestion control // IEEE/ACM Transactions on Networking. – 2000. – 8. – P. 556 – 567.
4. Paganini F., Doyle J.C., Low S.H. Scalable laws for stable network congestion control // Proc. of IEEE Conference on Decision and Control. – 2001. – 1. – P. 185 – 190.
5. Low S.H., Srikant R. A Mathematical Framework for Designing a Low-Loss, Low-Delay Internet // Network and Spatial Economics. – 2004. – 4 (1). – P. 75 – 102.
6. Altman E. et al. The evolution of transport protocols: An evolutionary game perspective //Computer Networks. – 2009. – Т. 53. – №. 10. – C. 1751-1759.
7. Smith, John Maynard. Evolution and the Theory of Games. Cambridge university press, 1982.
8. E. Altman, N. Bonneau, M. Debbah, G. Caire, An evolutionary game perspective to ALOHA with power control, in: Proceedings of the 19th International Teletraffic Congress, Beijing, 29 August–2 September, 2005.
9. Han, Zhu, et al. Game theory in wireless and communication networks. Cambridge University Press, 2012.
10. Meyn, Sean P. Control techniques for complex networks. Cambridge University Press, 2008.
11. Filippov, Aleksei Fedorovich, and Felix Medland Arscott, eds. Differential Equations with Discontinuous Righthand Sides: Control Systems. Vol. 18. Springer, 1988.
12. Border, Kim C. Fixed point theorems with applications to economics and game theory. Cambridge university press, 1989.